\documentstyle[11pt]{article}
\begin{document}
\begin{flushright}SJSU/TP-95-12\\July 1995\end{flushright}
\vspace{1.7in}
\begin{center}\Large{\bf How to Measure a Beable}\\
\vspace{1cm}
\normalsize\ J. Finkelstein\footnote[1]{
        Participating Guest, Lawrence Berkeley National Laboratory\\
        \hspace*{\parindent}\hspace*{1em}
        e-mail: FINKEL@theor3.lbl.gov}\\
        Department of Physics\\
        San Jos\'{e} State University\\San Jos\'{e}, CA 95192, U.S.A
\end{center}
\begin{abstract}
A brief discussion is given of measurement within the context of a
theory of ``beables'', e.g.\ theories of de Broglie, Bohm, Bell,
Vink, and also ``modal'' theories.  It is shown that even in an
ideal von Neumann measurement of a beable, the measured value may
not agree with the value which the beable had prior to the
measurement.
\end{abstract}
\newpage
\section{Introduction}
In standard quantum theory, the state of a system is described by
a state vector $|\psi \rangle $, an element of the Hilbert space of
possible states of the system.  There have been several suggestions
for ``completing'' this description by supposing that,
unlike in standard quantum theory, a system could
possess definite values of certain quantities even if the state
vector was not an eigenstate of the operators associated with those
quantities.  In the theory of de Broglie [1] and Bohm [2], particles
are taken to have, at each time, definite values of position.
Bell [3] has proposed a theory in which it is the fermion number at
(discretized) positions that has definite values.
Vink [4] has shown how the formulation given by Bell can apply to
any discrete quantity, and that in an appropriate limit this
formulation reproduces the causal theory of Bohm.
In the ``modal''
interpretations of van Fraassen [5], Kochen [6], Healey [7],
Dieks [8], and Bub [9], the identity of the quantities whose
values are definite can depend on the state vector, and so
can be different at different times.  In all of these theories
[1-9] that we are considering, the time dependence of the state
vector $|\psi \rangle $ is given by the Schr\"{o}dinger equation;
there is no ``collapse'' of the state vector.

Bell [3] has used the term ``beable'' to refer to a quantity whose
value can be said to actually exist, as opposed to an ``observable'',
which takes a value only when it is measured.  Bell's intention
was to have a theory in which the notion of measurement would not
be fundamental; nevertheless, it is important to understand how and
in what sense one could measure the value of a beable, and in this
paper we will attempt to contribute to that understanding.
In particular, we will discuss the question of whether the result
of measuring a beable will necessarily
correspond to the value that the beable had before the measurement.

First, let us define the notation we will use. We will
denote by $B$ a quantity which is a beable, and by $\hat{B}$ the
corresponding Hilbert-space operator, and let us expand the
state vector upon the eigenvectors of $\hat{B}$:
\begin{equation}
|\psi\rangle = \sum_{i}c_{i}|\psi _{i}\rangle
\end{equation}
where $\hat{B}|\psi_{i}\rangle = b_{i}|\psi _{i}\rangle $.
More generally, we can understand $B$ to represent a
collection of beables;
if we are considering the theory of Bohm, we should understand
the summation in eq.\  1 to represent an integral, and the states
$|\psi_{i}\rangle$ to be (unnormalized) position eigenstates.
We are assuming that the system whose state vector is written in
eq.\  1 does possess a definite value of $B$, which must equal
$b_{i}$ for some $i$; in the case in which the spectrum of
$\hat{B}$ is not degenerate, this corresponds to a single term
in the summation in eq.\  1.  We will denote by
$|\psi\rangle ^v$ the specification both of the
state vector and of the value of $B$, and we will indicate the value
of $B$ by placing a bar above the appropriate term in the expansion
of $|\psi\rangle$ upon the eigenvectors of $\hat{B}$.
Thus for example we would write
\begin{equation}
  |\psi\rangle ^{v} = c_{1}\overline{|\psi_{1}\rangle} +
  c_{2}|\psi_{2}\rangle
\end{equation}
in the case in which the state vector was
$c_{1}|\psi_{1}\rangle + c_{2}|\psi_{2}\rangle$,
and in which the value of $B$ was $b_1$.
It might be tempting to say that the bar in eq.\  2
indicates that the first term on the right-hand side is the
``correct'' one, but more precisely
eq.\  2 indicates that the state vector contains {\em both} terms,
and that the correct value of $B$ corresponds to the
first term.  We can use the notation introduced in eq.\  2 when
discussing modal interpretations also, if we understand that
$B$ represents a quantity which is picked out by the state vector
at the time at which eq.\  2 is supposed to apply.

To make a measurement upon the system, let us couple it to a
second system, which we shall call ``the apparatus'', and let the
apparatus initially have state vector $|A_{0}\rangle$. We will
represent the interaction between the system and the apparatus
by an arrow, and assume, as usual, that for each i
\begin{equation}
  |\psi_{i}\rangle|A_{0}\rangle\Longrightarrow
  |\psi_{i}\rangle|A_{i}\rangle
\end{equation}
with $\langle A_{i}|A_{j}\rangle =\delta_{ij}$.
Pauli has called such a measurement, which leaves
the system in the eigenstate corresponding to the measured value,
a ``measurement of the first kind''.  Instead, we will use the
term ``von Neumann measurement'' for any interaction between
system and apparatus that satisfies eq.\  3.  It of course follows
 from eq.\  3 that
\begin{equation}
  \sum_{i}c_{i}|\psi_{i}\rangle|A_{0}\rangle\Longrightarrow
  \sum_{i}c_{i}|\psi_{i}\rangle|A_{i}\rangle
\end{equation}
Since in the theories we are considering the state vector never
collapses, the ``result'' of the measurement is reflected
not in the final state vector, but instead in the final values
of the beables.  We will only consider cases in which each
$|A_{i}\rangle$ is an eigenstate of beable operators for
the apparatus; then, in
the notation we have introduced above, we
could indicate the result of the measurement by putting a bar
above the appropriate term on the right-hand side of eq.\  4.

We will not be discussing any measurement of a quantity which
is not a beable (e.g., a measurement of momentum for Bohm);
we will consider only measurements of quantities which have
definite values before the measurement.  We will call a
``faithful measurement'' one in which the measured value of
the beable is the same as the value before the measurement.
For example,
\begin{equation}
    [c_{1}\overline{|\psi_{1}\rangle}+c_{2}|\psi_{2}\rangle]
   \overline{|A_{0}\rangle}
   \Longrightarrow c_{1}\overline{|\psi_{1}\rangle
    |A_{1}\rangle}+c_{2}|\psi_{2}\rangle|A_{2}\rangle
 \end{equation}
represents a faithful measurement, but
\begin{equation}
    [c_{1}\overline{|\psi_{1}\rangle}+c_{2}|\psi_{2}\rangle]
  \overline{|A_{0}\rangle}
  \Longrightarrow c_{1}|\psi_{1}\rangle
 |A_{1}\rangle+c_{2}\overline{|\psi_{2}\rangle|A_{2}\rangle}
\end{equation}
represents a non-faithful measurement, in which before the
measurement the value of $B$ was $b_1$ but the result of
the measurement (as reflected in the final value of the
apparatus beables) was $b_2$.

Not every interaction that is a von Neumann
measurement is faithful; whether or not an interaction is
faithful can depend on the details of the Hamiltonian
responsible for the interaction, even if we consider only
those Hamiltonians which yield
von Neumann measurements. We will show this in the next section
by displaying two examples of Hamiltonians, both of
which yield von Neumann measurements (as defined by eq.\  3);
one of these Hamiltonians yields measurements which are always
faithful (as in eq.\  5), but the other Hamiltonian yields
measurements which are sometimes unfaithful (as in eq.\  6).
These examples assume the beable dynamics suggested by Bell [3]
and elaborated upon by Vink [4], but in the third section we
argue that with {\em any} beable dynamics
von Neumann measurements will not
necessarily be faithful.

\section{Faithful and unfaithful examples}
In the beable dynamics suggested by Bell [3], and elaborated upon
by Vink [4], the probability that the value of the beable jumps
 from $b_i$ to $b_j$ in time $dt$ is denoted by $T_{ij}dt$, and is
given by
\begin{equation}
    T_{ij}=\mbox{max}[-2\, \mbox{Im}
    ({\textstyle\frac{c_{i}}{c_{j}}}H_{ji})\; ,\; 0\; ],
\end{equation}
where $H_{ji}$ is the matrix element of the Hamiltonian and $c_i$
is defined in eq.\ 1.  It follows from eq.\ 7 that the value of
the beable $B$ will be constant if $H$ commutes with $\hat{B}$.
This dynamics is of course stochastic;
however, the examples we consider below will be determinate.
To avoid confusion, we should note that we are not now
considering the alternatives to eq.\ 7 suggested by Vink
(eqs.\ 15 and 16 of ref.\ 4), nor will we be concerned with the
further suggestion made by Vink that (in spite of the theorem of
Kochen and Specker [10]) {\em all} quantities could be taken
to be beables.  We will only be interested in those beables
which are, in fact, measured.

We want to consider a system (for which we use a superscript s)
which is interacting with an apparatus (superscript a).
We model each of them as a spin-$\frac{1}{2}$ particle, and
take the beables to be the z components of the spin of each.
Thus the basis of the combined system which corresponds to the
$|\psi_{i}\rangle$ in eq.\ 1 will be the product basis,
consisting of the four elements $|+\rangle^{s}|
+\rangle^{a},\; |+\rangle^{s}|-\rangle^{a},\;
|-\rangle^{s}|+\rangle^{a}$ and $|-\rangle^{s}
|-\rangle^{a}$. Let $|\Psi\rangle$ denote the state vector for the
combined system, and say that before the measurement we have
\begin{equation}
  |\Psi\rangle_{t=0}=[c_{+}|+\rangle^{s}+c_{-}
  |-\rangle^{s}]|+\rangle^{a}.
\end{equation}

\underline{First example: a faithful measurement}. Let the
interaction between $s$ and $a$ take place between times $t=0$
and $t=\tau$, and take the  Hamiltonian to be,
during the interaction,
\begin{equation}
   H={\textstyle\frac{\pi}{4\tau}}
   (I-\sigma_{z})^{s}\otimes\sigma_{y}^{a}.
\end{equation}
Note that we are taking this $H$
 to be the {\em full} Hamiltonian for the
entire duration of the interaction. Then
\begin{eqnarray}
   U(\tau,0)&=&exp(-iH\tau)\nonumber \\
   &=&{\textstyle\frac{1}{2}}((I+\sigma_{z})^{s}\otimes I^{a}
       -i(I-\sigma_{z})^{s}\otimes \sigma_{y}^{a}).
\end{eqnarray}
 From eqs.\ 8 and 10, we see that after the interaction we have
\begin{equation}
  |\Psi\rangle_{t=\tau}=c_{+}|+\rangle^{s}|+\rangle^{a}
  +c_{-}|-\rangle^{s}|-\rangle^{a}.
\end{equation}
Then if, in eqs.\ 8 and 11, we take alternatively $c_{+}=1,\;
c_{-}=0$ and $c_{+}=0,\; c_{-}=1$, we can see that this interaction
is indeed a von Neumann measurement (as in eq.\ 3),
with the identifications $|A_{0}\rangle = |A_{+}\rangle =
|+\rangle^{a}$ and $|A_{-}\rangle =|-\rangle^{a}$.

Since $H$ commutes with $\sigma_{z}^s$,
 the value of the
beable $S_{z}^s$ will not change during the interaction;
this means that we have a faithful
measurement.  In the notation introduced in eq.\ 2, suppose that
before the interaction the bar belongs over the first term on
the right-hand side of eq.\ 8; that is, suppose
\begin{equation}
|\Psi\rangle^{v}_{t=0}= c_{+}\overline{|+\rangle^{s}|+\rangle^{a}}
     +c_{-}|-\rangle^{s}|+\rangle^{a}.
\end{equation}
Then, since after the interaction the bar must still be over the
state $|+\rangle^{s}$, and since that state only appears once on the
right-hand side of eq.\ 11, we know that we have
\begin{equation}
  |\Psi\rangle^{v}_{t=\tau}=
  c_{+}\overline{|+\rangle^{s}|+\rangle^{a}}
     +c_{-}|-\rangle^{s}|-\rangle^{a}.
\end{equation}
Eqs.\ 12 and 13 (together with the analogous equations with the
bar starting over the second term in eq.\ 8) tell us that we have
a faithful measurement. In fact it is an obvious generalization
of this example that, given the beable dynamics defined by eq.\ 7,
any  Hamiltonian which commutes with the measured beable,
and which yields a von Neumann measurement, will yield a
faithful measurement. This will be the case with the usual
prescription, which is to take the  Hamiltonian during the
interaction to
be proportional to the operator representing the measured
quantity. However, this prescription is not included in
the requirement
 (eq.\ 3) which we have used to define a von Neumann measurement.

\underline{Second example: an unfaithful measurement}.
This time let the interaction occur between $t=0$ and $t=4\tau$,
during which time the Hamiltonian is (with $\hbar=1$)
\begin{equation}
H=\left\{ \begin{array}{ll}-\frac{\pi}{4\tau}
  \sigma_{z}^{s}\otimes\sigma_{y}^{a} &
  \mbox{for $0<t<\tau$} \\
  -\frac{\pi}{2\tau}\sigma_{x}^{s}\otimes I^{a}
  &\mbox{for $\tau<t<2\tau$} \\
  +\frac{\pi}{4\tau}I^{s}\otimes\sigma_{y}^{a}
  &\mbox{for $2\tau<t<3\tau$} \\
  +\frac{\pi}{2\tau}\sigma_{x}^{s}\otimes I^{a}
  &\mbox{for $3\tau<t<4\tau$}
  \end{array} \right.
\end{equation}
Then
\begin{eqnarray}
  U(\tau,0)&=&{\textstyle \frac{1}{\sqrt{2}}}(I^{s}\otimes I^{a}
  +i\sigma_{z}^{s}\otimes \sigma_{y}^{a})  \\
  U(2\tau,\tau) &=& i\sigma_{x}^{s}\otimes I^{a} \\
  U(3\tau,2\tau) &=&{\textstyle \frac{1}{\sqrt{2}}}I^{s}
  \otimes (I-i\sigma_{y})^{a} \\
  U(4\tau,3\tau) &=& -i\sigma_{x}^{s}\otimes I^{a},
\end{eqnarray}
and then
\begin{eqnarray}
  U(4\tau,0) &=& U(4\tau,3\tau)U(3\tau,2\tau)
  U(2\tau,\tau)U(\tau,0)\nonumber\\
   &=&{\textstyle\frac{1}{2}}((I+\sigma_{z})^{s}\otimes I^{a}
       -i(I-\sigma_{z})^{s}\otimes \sigma_{y}^{a}).
\end{eqnarray}
By comparing eqs.\ 10 and 19 , we can see that this second
example has given us the same von Neumann measurement as did
the first.

Now let us take the initial state to be given by eq.\ 8
with the specification $c_{+}=c_{-}=\frac{1}{\sqrt{2}}$.
Let us also take the initial value of the beable $S_{z}^{s}$
to be $+$ (the initial value of $S_{z}^{a}$ is necessarily $+$);
then we can write
\begin{equation}
   |\Psi\rangle^{v}_{t=0}={\textstyle \frac{1}{\sqrt{2}}}
   \overline{|+\rangle^{s}|+\rangle^{a}}
   +{\textstyle \frac{1}{\sqrt{2}}}
   |-\rangle^{s}|+\rangle^{a}.
\end{equation}
Now from eq.\ 15 we see that
\begin{equation}
   |\Psi\rangle_{t=\tau}={\textstyle \frac{1}{2}}
   |+\rangle^{s}|+\rangle^{a}
   -{\textstyle \frac{1}{2}}|+\rangle^{s}|-\rangle^{a}
   +{\textstyle \frac{1}{2}}|-\rangle^{s}|+\rangle^{a}
   +{\textstyle \frac{1}{2}}|-\rangle^{s}|-\rangle^{a}.
\end{equation}
Since for $0<t<\tau$, $H$ commutes with $\sigma_{z}^s$,
the value
of $S_{z}^s$ is still $+$ at $t=\tau$, and so in eq.\ 21 the bar
must go over one of the first two terms on the right-hand side.
Because of the stochastic nature of the dynamics, we cannot say
with certainty which one (in fact these two alternatives have
equal probabilities); this ambiguity will be irrelevant when
we get to the end of the calculation.  So let's choose to put
the bar over the first term; this gives
\begin{equation}
   |\Psi\rangle^{v}_{t=\tau}={\textstyle \frac{1}{2}}
   \overline{|+\rangle^{s}|+\rangle^{a}}
   -{\textstyle \frac{1}{2}}|+\rangle^{s}|-\rangle^{a}
   +{\textstyle \frac{1}{2}}|-\rangle^{s}|+\rangle^{a}
   +{\textstyle \frac{1}{2}}|-\rangle^{s}|-\rangle^{a}.
\end{equation}

For $\tau<t<2\tau$, eq.\ 7 implies that the values of the beables
will not change.  To see this, note that the solution of the
Schr\"{o}dinger equation in this time interval is
\begin{eqnarray}
  |\Psi\rangle_{t}&=&
  {\textstyle \frac{1}{2}}[|+\rangle^{s}+|-\rangle^{s}]
  |+\rangle^{a}\exp [i\pi (t-\tau)/(2\tau)]\nonumber \\ & &
  \mbox{}-{\textstyle \frac{1}{2}}[|+\rangle^{s}-|-\rangle^{s}]
  |-\rangle^{a}\exp [-i\pi (t-\tau)/(2\tau)].
\end{eqnarray}
All matrix elements of $H$ are real, and from eq.\ 23, the ratio
$c_{i}/c_{j}$ that appears in eq.\ 7 has
the value one in all cases
in which $H_{ji}\neq 0$.  Thus $(c_{i}/c_{j})H_{ji}$ is in all
cases real, and so eq.\ 7 implies that all $T_{ij}$ vanish.
So the values of $S_{z}^{s}$ and $S_{z}^a$ at $t=2\tau$ are
the same as they were at $t=\tau$, namely both are $+$. Then
using eqs.\ 16 and 21 we have
\begin{equation}
  |\Psi\rangle^{v}_{t=2\tau}=
  {\textstyle \frac{i}{2}}|-\rangle^{s}|+\rangle^{a}
  -{\textstyle \frac{i}{2}}|-\rangle^{s}|-\rangle^{a}
  +{\textstyle \frac{i}{2}}\overline{|+\rangle^{s}|+\rangle^{a}}
  +{\textstyle \frac{i}{2}}|+\rangle^{s}|-\rangle^{a}.
\end{equation}

For $2\tau<t<3\tau$, $H$ commutes with $\sigma_{z}^s$,
so the value
of $S_{z}^{s}$ remains $+$; then from eq.\ 17 we have
\begin{equation}
  |\Psi\rangle^{v}_{t=3\tau}={\textstyle \frac{i}{\sqrt{2}}}
  |-\rangle^{s}|+\rangle^{a}+{\textstyle \frac{i}{\sqrt{2}}}
  \overline{|+\rangle^{s}|-\rangle^{a}}.
\end{equation}
If in eq.\ 22 we had chosen to put the bar above the second term
instead of the first, we would have obtained eq.\ 25 exactly
as above.  Finally, for $3\tau<t<4\tau$, $H$ commutes with
$\sigma_{z}^a$,
so the value of $S_{z}^{a}$ has the same value at
$t=4\tau$ as it did at $t=3\tau$, namely $-$; then from eq.\ 18
we see
\begin{equation}
  |\Psi\rangle^{v}_{t=4\tau}={\textstyle \frac{1}{\sqrt{2}}}
  |+\rangle^{s}|+\rangle^{a}+{\textstyle \frac{1}{\sqrt{2}}}
  \overline{|-\rangle^{s}|-\rangle^{a}}.
\end{equation}
By comparing eqs.\ 20 and 26, we see that this measurement is
not faithful.

What is most surprising about this example is not that the
final and initial values of $S_{z}^{s}$ are different, but
rather that the measured value of  $S_{z}^{s}$
{\em as read from the apparatus} differs from the initial
value. In fact, the measured value (namely $-$) differs
 from the value that  $S_{z}^{s}$ had {\em for the entire
time when $s$ and $a$ were actually interacting}.
To see this, note that, from eq.\ 7, $s$ and $a$ evolve
independently for $t>\tau$, and that  $S_{z}^{s}$ has the
value $+$ for $0<t<\tau$. In the terminology we are using,
the ``measurement'' is not completed until $t=4\tau$,
when the state vector has the form given in eq.\ 4,
with each $|A_{i}\rangle$ an eigenstate of apparatus beables.

We have seen that the Hamiltonians in our two examples,
although they yield identical von Neumann measurements,
treat the values of the beables differently.  The Hamiltonian
in the first example yields a faithful measurement, for any
values of $c_+$ and $c_-$ in eq.\ 8.  The Hamiltonian in the
second example would yield a faithful measurement for
$c_{+}=1,\; c_{-}=0$ or for $c_{+}=0,\; c_{-}=1$
(as is required for any von Neumann measurement), but for
$c_{+}=c_{-}=\frac{1}{\sqrt{2}}$ it yields an unfaithful
measurement.

\section{Discussion}
The second example of the preceding section shows that,
assuming Bell-Vink dynamics for discrete beables (eq.\ 7),
not every von Neumann measurement is faithful. In that example,
the measured value of $S_{z}^{s}$ differs from the
value that  $S_{z}^{s}$ had {\em for the entire
time when $s$ and $a$ were actually interacting}. Could this
example be somehow exceptional?  In the example, we took
$c_+$ and $c_-$ to be precisely equal.  However, this choice
was made to get a simple example in which, in spite of the
stochastic nature of the dynamics, the measurement would always
give the ``wrong'' answer.  With other choices of $c_+$ and
$c_-$, the answer would be wrong some of the time.  It is
complete faithfulness, not occasional unfaithfulness,
which is the exceptional case.

Since Vink [4] has shown that, in an appropriate limit, this
dynamics reproduces the causal theory of Bohm, one might
therefore suspect that a von Neumann measurement of position
in Bohm's theory might not correspond to the particle's
position before the measurement.  That this suspicion is
correct can be seen from the work of Englert, Scully, Sussman
and Walther [11]. The example discussed in ref.\ 11, although
not presented in quite the terms we are using here, can easily
be modified to be an example of an unfaithful von Neumann
measurement of position in Bohm's theory. (In fact, the second
example presented in this paper could be considered as a
discretized version of the example discussed in ref.\ 11.)
Of course if one does make the usual choice
(see for example p.\ 109 of ref.\ 12)
of a Hamiltonian which is proportional to the position
operator of the measured particle, one will be assured of
obtaining a faithful measurement of position in Bohm's theory.
In the
terminology we are using here, the measurement is not over
until the states $|A_{i}\rangle$ are eigenstates of apparatus
beables.  In the example discussed in ref.\ 11 (see also
Dewdney, Hardy, and Squires [13]) this does not happen until
the measured particle has moved away from the apparatus,
and so the (full) Hamiltonian is {\em not} proportional
to the position operator of the particle for the entire
duration of the ``measurement''.

Vink [4] has also suggested alternative dynamics to the one
suggested by Bell [3].  However, it is very difficult to
see how {\em any} dynamics for
discrete beables can avoid all examples of the type we have
considered above, without putting restrictions on the
Hamiltonian.  The difficulty can be seen even without
considering a measurement situation.  So let's
temporarily forget the apparatus, and say that the system
$s$ evolves for some time with a Hamiltonian
$H=\frac{\pi }{4\tau}\sigma_{y}^{s}$.
If at $t=0$ the state vector is
\mbox{$|\psi\rangle_{t=0}=\frac{1}{\sqrt{2}}
(|+\rangle^{s}+|-\rangle^{s})$,}
at $t=\tau$ it will be
$|\psi\rangle_{t=\tau}=|-\rangle^{s}$.
Thus at $t=\tau$ the value of $S_{z}^s$ will surely
\mbox{be $-$}
(since there is no other term in $|\psi\rangle$),
irrespective of its value at $t=0$. So at $t=\tau$ the system
has ``forgotten'' the value which $S_{z}^s$ had at $t=0$.
Unless the beable dynamics were non-Markovian, there could
not therefore be any correlation between the values of
$S_{z}^s$ at $t=0$ and $t=8\tau$, even though with this
Hamiltonian we have $U(8\tau,0)=I$. And if this kind of
forgetfulness were to happen to our system in the middle
of a measurement, there would be no relationship between
the final (i.e., the measured) and the initial values of
the beable.

We have been assuming that we are dealing with beables whose
identity is independent of time, but that is not the case in
modal theories [5-9]. Therefore it is not ruled out that
there could be  modal dynamics which would
guarantee that any von Neumann measurement be faithful.
However, this does not seem likely, since one would have
thought that the necessity of taking different bases in
eq.\ 1 at different times would make it harder, not easier,
to guarantee faithfulness.

What we have been calling a von Neumann measurement (eq.\ 3)
is certainly not the most general interaction that should
be considered to be a measurement; a ``measurement of the
second kind'' can change the state of the system even if
the system started out in an eigenstate of the quantity being
measured. On the other hand, from the point of view of this
paper, a von Neumann measurement is also not sufficiently
restrictive, since it does not ensure faithfulness. For a
von Neumann measurement with Bell-Vink (or its limiting
case Bohmian) dynamics,
it is at least possible to formulate
a restriction on the Hamiltonian  which {\em does} ensure
that the measurement be faithful, namely that the Hamiltonian
commute with the beable to be measured.  Obviously this
particular restriction could not be applied to a measurement
of the second kind.

\vspace{1cm}
\begin{flushleft}
\large {\bf Acknowledgements}
\end{flushleft} \normalsize
\vspace{0.5cm}
I would like to acknowledge very helpful conversations with
Andy Elby, as well as the hospitality of the
Lawrence Berkeley National Laboratory.

\end{document}